# Dielectrophoresis-Enhanced Graphene Field-Effect Transistors for Nano-Analyte Sensing

Nezhueyotl Izquierdo,* Ruixue Li,* Peter R. Christenson, Sang-Hyun Oh, and Steven J. Koester

*Department of Electrical and Computer Engineering, University of Minnesota, 200 Union Street SE, Minneapolis, MN 55455, United States*

## Abstract

Dielectrophoretic (DEP) sensing is an extremely important sensing modality that enables the rapid capture and detection of polarizable particles of nano-scale size. This makes it a versatile tool for applications in medical diagnostics, environmental monitoring, and materials science. Because DEP relies upon the creation of sharp electrode edges, its sensitivity is fundamentally limited by the electrode thickness. Graphene, with its monolayer thickness, enables scaling of the DEP force, allowing trapping of particles at graphene edges at ultra-low voltages. However, to date, this enhanced trapping efficiency of graphene has not been translated into an effective sensing geometry. Here, we demonstrate the expansion of graphene DEP trapping capability into a graphene field effect transistor (GFET) geometry that allows the trapped particles to be electrically detected. This four-terminal multi-functional hybrid device structure operates in three distinct modes: DEP, GFET, and DEP-GFET. By segmenting the channel of the GFET into multiple parallel channels, greatly increased density of particle trapping is demonstrated using fluorescence microscopy analysis. We show further enhancement of the trapping efficiency using engineered "nano-sites," which are holes in the graphene with size on the order of 200-300 nm. Scanning electron microscope analysis of immobilized gold nanoparticles (AuNPs) shows trapping efficiency >90% for properly engineered nano-sites. Use nano-site trapping, we also demonstrate real-time, rapid electrical sensing of AuNPs, with >2% current change occurring in 4.1 seconds, as well as rapid sensing of a variety of biomolecule-coated nanoparticles. This work shows that graphene DEP is an effective platform for nanoparticle and bio-molecule sensing that overcomes diffusion-limited and Brownian motion-based interactions.

*Equal contributors

As a two-dimensional (2D) material, graphene possesses favorable properties as a biosensor transducer channel material,[1] including high mobility,[2] chemical and mechanical stability, biocompatibility, and ease of surface functionalization.[3] In a liquid medium, graphene field-effect transistor (GFET) sensors have the potential for fast response due to their strong surface sensitivity.[4] However, traditional GFET sensors often rely upon diffusion and Brownian motion to produce a recognition event. Therefore, the detection of target particles, particularly for solutions approaching the limit of detection (LOD), can still require a long incubation time in order to observe a measurable response.[5-8] The integration of dielectrophoresis (DEP)[9] within the device structure could offer a promising method to speed up the response time of GFET sensors by overcoming the Brownian motion to concentrate target particles locally through an attractive force.[10] DEP is a well-established technique and the immobilization of various biomolecules and particles, including DNA,[11] enzymes,[12] polystyrene (PS) beads,[13,14] and gold nanoparticles (AuNPs)[15] have been demonstrated. Furthermore, the use of DEP to enhance sensitivity and reduce response times is a common practice in many sensor platforms, as reviewed by Henriksson, *et al.*,[16] who described how DEP improves mass transfer, a key bottleneck in traditional biosensor geometries. As an example, Sharma, *et al.*, integrated DEP into a photonic biosensor, improving the response time from 60 minutes to 1 minute, with improved sensitivity.[17] In another example, Szymborski, *et al.*, developed a DEP-based surface-enhanced Raman scattering (SERS) sensor for detecting cancer cells in microfluidic chips.[18] DEP was shown to improve the sensor limit of detection (LOD) down to 20 cells/mL, with a response time of only 7 minutes.

Despite the benefit of DEP for improving sensor operation, its use in conjunction with GFET sensors is very limited in the literature. Most notably, Kumar, *et al.* used DEP to enhance GFET sensor operation and detect antibiotic-resistant bacteria at the single-cell level.[19] Their DEP-assisted GFETs achieved a response time of approximately 5 minutes, a 9-fold reduction in response time, demonstrating the effectiveness of DEP in accelerating sensor performance. However, the approach used by Kumar *et al.* was limited in that the DEP and sensing were conducted sequentially, rather than simultaneously, by applying a series of DC biases to the liquid gate itself, and then measuring the FET. While straightforward, this sequential approach limited the response time improvement that could be achieved.

To overcome the limitations of previous DEP GFET sensors, the inherent properties of graphene can be utilized. In particular, Barik *et al.* showed that the DEP force is especially strong at the atomically-thin edges of graphene,[20] and is capable of efficiently trapping both PS beads and biologically relevant species such as DNA at ultra-low voltages. However, while that report demonstrated the potential of graphene for use in DEP trapping, the device structure was not well-suited for use in sensors, since the trapping only occurred only to the periphery of the device. Furthermore, a two-terminal "varactor" structure[21,22] was used, which is not ideal for biosensing using DEP, since the AC bias would have to be applied to the same buried electrode needed to interrogate the sensor, making real-time readout difficult.

In this work, we overcome the limits of this prior work by proposing and demonstrating a novel four-terminal DEP-GFET geometry. This device separates the buried and liquid gate electrodes, allowing the sensor readout and DEP force to be applied independently and potentially, simultaneously. It also uses a GFET geometry for sensor readout, where the graphene can be patterned into smaller segments that provides high sensitivity by providing a large area of the sensor that interacts with the trapped particles. In this way, it meets the key metrics needed for an effective DEP-enhanced sensor: high trapping efficiency, high sensitivity, and fast response time.

**Device Structure and Fabrication.** Our novel sensor design is shown in Figure 1a, where the complete fabrication process flow is provided in the Supporting Information (Figure S1). The device structure consists of four terminals: a buried local back gate (LBG) is used for controlling DEP trapping, and a liquid gate (LG) electrode modulates the GFET sensor. The source and drain electrodes are located on opposite sides of the LBG. A top view of a completed fabricated device is shown in Figure 1b.

**Modes of Operation.** The four-terminal geometry allows the sensor to be operated in one of three modes: GFET only, DEP only, and DEP-GFET, as illustrated in Figure 2. In this section, we briefly describe these different modes of operation, where more details are provided in the Supporting Information. In the "GFET" sensing mode, DEP is not used and the device operates as a conventional GFET sensor. In this configuration (Figure 2a), the sensor tracks the Dirac point, $V_{Dirac}$, and/or the drain current, $I_D$, response to the analyte concentration. In the second "DEP" mode of operation (Figure 2b), DEP is used to trap particles at the graphene edges, and fluorescence microscopy is used to identify trapping at the graphene edges. A final sensing mode is the "DEP-GFET" configuration, depicted in Figure 2c, where the two sensing configurations are combined. This mode has the advantage of separating DEP and sensing sweeps, potentially allowing the trapping to DEP bias to be applied during sensing. Additional details on the specific sensing modes enabled by this configuration are provided below. The measurement setup is also described in Figure S2.

**Fluorescence Measurements of PS Beads in DEP Mode.** As an initial step to characterize the DEP trapping efficiency, we analyzed the device operation using fluorescence (FL) microscopy (Figure 3). As depicted schematically in Figures 3a and 3b, a localized increase in the concentration of fluorophore-decorated particles resulting from the DEP force should produce FL intensity bands at the trap sites located at the edges of the graphene. Here, fluorophore-decorated PS particles (42 nm, Bangs Laboratories, Inc) were selected to evaluate the trapping capabilities for the device structure. The experiments are performed in 0.01X phosphate buffered saline (PBS) solution to minimize Joule heating and ionic shielding effects.[23] For the DEP-off state (1 mV$_{PP}$, 800 kHz) trap sites are vacant, evidenced by the FL image and the FL 3D surface plot map (Figure 3c). In comparison, the DEP-on state (1.3 V$_{PP}$, 800 kHz) results in essentially full trap site occupancy (~99%) (Figure 3d). Thus, the multi-channel graphene array geometry is effective, with the trap site length increased by five-fold compared to the situation that would have occurred without the segmented graphene channel. An FL intensity line profile along the length of the LBG electrode is provided in Figure 3e. A comparison of the DEP-off and DEP-on states shows an increase in the FL intensity at all ten graphene channel edge positions. In Figure 3f, three sequential DEP (on-off) pulses are measured by monitoring the mean FL intensity at all trap sites (S Video 1). After baseline correction, the maximum FL intensity value during DEP-on is 55× the calculated baseline (DEP-off) standard deviation, and the FL intensity increases to 50% of the maximum fluorescence in only 4.5 seconds, and no permanent trapping of the PS beads is observed. We also evaluated the effect of increasing $V_{PP}$ on the FL intensity. A series of on-off DEP pulses with sequential 100 mV$_{PP}$ increases was performed and the results are presented in Figure 3g. The normalized FL intensity increases in response to increasing the trapping bias, as expected. Additional FL trapping results are shown in S. Video 2.

    The fluorescence data clearly show that diffusion alone results in minimal analyte accumulation at the graphene surface, evidenced by a very weak fluorescence response. This indicates that, under natural diffusion, target analytes do not reach the graphene surface in sufficient concentration to produce a strong fluorescence signal. However, upon applying DEP, a

significant fluorescence signal increase is observed, demonstrating a rapid and targeted concentration of analytes at the sensor surface. This sharp increase confirms that DEP effectively improves the limitations of passive diffusion, actively driving analytes to the graphene and enabling a much stronger fluorescence response.

**Fluorescence and SEM of AuNPs in DEP Mode.** In order to precisely count the number of trapped particles and determine the precise location of the graphene participating in trapping along the channel edge, we performed additional experiments using gold nanoparticles (AuNPs). AuNPs can be imaged in a scanning electron microscope (SEM), yet are also biologically relevant with applications in bio-medicine and sensing. The AuNPs had 150-nm diameter and were trapped using DEP at a frequency of 1.5 MHz and a LBG bias of 2 $V_{PP}$. In this regime, permanent trapping occurred, meaning that after removing the DEP excitation, the particles remained attached to the graphene edges through van der Waals forces, which allowed for subsequent SEM imaging. In Figure 4a, a comparison of DEP trapping (FL) and post-DEP immobilization (BF) is shown. SEM characterization (Figures 4b and 4c) of immobilized AuNPs provides a simple method to evaluate trap sites with high spatial resolution and allow particle counting and trapping morphology analysis.

From the SEM images, it can be seen that the AuNPs form a linear chain at the engineered trap site locations. In Figure 4b, approximately 60 AuNPs are seen to be trapped along the ~7.5 μm trap site length, and the strong clustering just at the graphene edge is a good indication of the locality of the DEP force. However, while the segmentation of graphene enhances the DEP interaction area of the graphene sensor, further improvement is needed. Therefore, we evaluated the use of electron-beam lithography to create high-density nanoscale trap sites or "nano-sites" within the graphene channel, with the intent to have a similar trapping effect as metallic nanopores used in previous work,[24,25] except with greatly simplified fabrication.

The nano-sites consisted of etched "holes" in the graphene created using a short oxygen plasma, where the fabrication details are described in the Supporting Information. As shown from the FL microscopy images in Figure 4d and S. Video 3, trapping using nano-sites spelling out the letters U-M-N were used, was clearly effective. An example schematic of the complex nano-site pattern design is presented in Figure 4e, and the specific location of the trapping is shown in Figure 4f. The regions between the nano-sites are less preferred than the edges that directly access the wider graphene regions, and this can be traced to series resistance effects from the graphene, as the AuNPs prefer to trap at sites where the resistance relative to the contact electrode is minimized. Numerical simulations (Figure 4g) further confirm that the nano-sites also have a strong electric field gradient which accounts for their excellent trapping behavior.

We also investigated the effect of the nano-site dimensions. Here, the hole spacing, $H_S$, was fixed while the hole length, $H_L$, was varied. We observed that increasing $H_L$ was found to increase the number of trapped AuNPs, resulting in an overall increase in the FL intensity. A line profile comparison of edge trapping and edge plus nano-site trapping shows a significant boost in normalized FL signal within the graphene channel (Figure 4h). The permanent trapping occupancy (PTO), defined in the Supporting Information, was measured for nano-sites with $H_L$ = 100, 300, and 500 nm, and it was found that PTO $\geq$ 90% for $H_L > 300$ nm, showing that the nano-sites must be larger than the AuNPs in order to trap effectively. Figure 4i further shows that more particles are trapped per nano-site as $H_L$ is increased. Additional analysis of the trapping statistics are shown in Figure S3.

**DEP-GFET Sensing of AuNPs.** In the next aspect of our work, we performed measurements in the DEP-GFET sensing mode, where the GFET and DEP measurements are combined. In order to perform these measurements in a controllable manner, care must be taken to avoid measurement artifacts. Therefore, we established a three-step DEP-GFET sensing protocol, and this is depicted in Figure 5a and described in detail in the Supporting Information. In brief, steps 1 and 3 are used to measure the $V_{Dirac}$ shift before and after applying DEP as an indirect sensing method. However, this method is only responsive to immobilized particles. Alternatively, in step 2, the response to the DEP excitation can be measured by monitoring $I_D$, allowing for real-time sensing.

Our initial measurements in DEP-GFET mode used fluorophore-decorated-AuNPs (100 pM, AuNP in 0.01X PBS) to evaluate the sensing protocol, and the results were compared with 0.01X PBS controls. The first sensing method utilized $V_{Dirac}$ tracking. A comparison of the relative $V_{Dirac}$ shift, after three sequential 2-minute DEP pulses, shows that a response only occurs in the presence of the active sample. Specifically, after an aggregate exposure of 6 minutes, the control sample produced only a -2 mV shift in $V_{Dirac}$, (Figure 5b), while the sample of 100 pM AuNPs produced a -40 mV shift (Figure 5c). This shift relative to the control is consistent over multiple trials on the same sample and the statistical results are shown in Figure 5d.

The second tracking method used was to monitor $I_D$ during application of the DEP force. Measuring the sensor by monitoring $I_D$ allows the response to be observed nearly instantaneously, and much faster than the $V_{Dirac}$ mode, which requires a baseline and response transfer curve measurement, reducing the granularity of the time response, adding several minutes to the total sensor response time beyond that needed for incubation. Before performing the real-time measurements, $I_D$-$V_G$ curves were measured in buffer solution with no particles, to ensure the GFET produced reasonable characteristic while the DEP force is being applied. Since the DEP bias is an AC signal, it only disturbs the $I_D$-$V_G$ characteristics by smearing the curve near the Dirac point, an effect that increases with increasing $V_{PP}$. These results are shown in Figure S4 and indicate that while the current modulation is reduced, a Dirac point can still be observed even at $V_{PP} = 2.5$ V. Using this measurement configuration, the DEP-$I_D$ response was measured for AuNPs and the results are shown in Figure 5e. Here, DEP-$I_D$ sensing of AuNPs is shown using both the electrical (red) and FL (black) response. The current response is almost immediate and changes by 2% in only 4.1 seconds, while the FL response takes much more time to build up, highlighting the benefit of the DEP-$I_D$ sensing mode.

**DEP-GFET Sensing of Biomolecules.** In the final aspect of our work, we used the DEP-GFET sensing scheme (in $I_D$ mode) to detect various biologically relevant classes of targets, including double-stranded DNA (ds-DNA), PS latex-beads coated with streptavidin protein, and a virus-like-particle (VLP). In all measurements described below and shown in Figure 5f, $I_D$ was measured for 10 minutes and normalized to the starting DEP-$I_D$ value. For the control measurements, a baseline in 0.01X PBS buffer was established. Here, the current was virtually unchanged ($I_D$ increased by $2 \pm 1$% over 7 trials). Details of the particle preparation and measurement conditions are provided in the Supporting Information. As shown in Figure 5f, ds-DNA produced an $I_D$ change of $-24 \pm 3$% (over 7 trials), while the streptavidin-coated PS beads produced an average $I_D$ change of $-44\% \pm 8$% (over 4 trials). Finally, the VLPs produced the largest change of $-60 \pm 9$% (over 4 trials). FL imaging of VLP trapping is also shown in Figure S5, where it can be seen that VLPs are isolated to the DEP trapping regions, as expected.

For all the measurements above, the reproducibility of the response was tested before and after DEP, to ensure that the change in current was not due to sample degradation or other non-sensing-related effect. Multiple DEP pulses in succession were analyzed to assess the reproducibility of the response, as well as potential graphene channel degradation. The maximum DEP-$I_D$ does not decrease after sequential DEP pulses, implying that graphene is not damaged during the application of DEP. Furthermore, SEM images of graphene-edge trap sites (Figure 4b) and/or nano-sites (Figure 4f) do not show graphene damage provided that the DEP excitation amplitude remained in the range of 0–2.5 $V_{PP}$.

The results demonstrate that the observed GFET electrical response is due to DEP-enhanced analyte accumulation at the graphene surface. The DEP-GFETs are already incubated in the target analyte solution when the DEP force is applied. The electrical response is therefore clearly due to the DEP-enhanced effect, and cannot stem from adding the target solution, which may change the ionic strength, or sample preparation byproducts, as the measurements before and after DEP-force are applied were taken in a consistent solution. Furthermore, for each sample condition—Streptavidin, VLP, and DNA—we fitted an exponential decay trend line to the current data over the first 30 seconds to model early response behavior under dielectrophoresis (DEP) influence. Similar to the AuNP results described above, from these fitted trends, we identified the earliest time at which each condition's current reached the 2% threshold. Streptavidin (10 pM), VLP (5 fM), and ds-DNA (1 nM) each demonstrated rapid response, reaching the 2% threshold within 3.9, 9.4, and 14.5 seconds, respectively, indicating graphene-edge DEP is effective for a wide range of targets in attracting analytes to the sensor surface. These DEP-assisted GFET measurements compare favorably with the even the best response times observed in traditional GFET biosensors upon analyte addition.[26,27]

**Limitations and Methods for Improvement.** While we have demonstrated the DEP-GFET operation for a variety of nanoparticles, a current limitation of our work is that the interaction is largely non-specific. It is well-established that GFET-based biosensors can provide selectivity by incorporating surface functionalization.[28-30] Therefore, a next step would be to determine the degree to which surface functionalization can be combined with DEP to enable selectivity. DEP could also be used in other ways to impart selectivity. For instance, while we focused on sensing target analytes trapped by positive DEP, negative DEP could provide a mechanism to reduce nonspecific adsorption, a persistent challenge for all sensor technologies. By removing unwanted nonspecific interactions at the surface of the graphene channel, the engineered probe-target response is isolated, improving the specificity and selectivity. Creating an array of multiplexed DEP-GFET devices working in unison could also provide pseudo-selectivity simply by operating individual sensors at different frequencies. Another interesting application of our DEP technique would be to utilize it to perform surface functionalization itself. Several groups have published results where AuNPs on graphene were used as functionalization elements. However, such techniques can damage the graphene, produce non-uniform coverage, and require long incubation times (e.g., 30 min).[31,32] Therefore, our DEP technique could be used as a simpler means of AuNP functionalization, where the particle position can be controlled precisely.

In conclusion, we have demonstrated a multi-modal four-terminal DEP-GFET device design that can effectively trap and sense a variety of nano-scale particles. Using FL microscopy, we provided evidence of improved DEP efficiency at engineered trap sites, located at the graphene-edge intersections with a LBG electrode. Immobilization of AuNPs at the trap sites further provided a technique to determine trapping location with precision, number of particles, and trapping morphology. Exposed graphene edges can be significantly increased by nano-site patterning to form complex and dense nano-site arrays of different dimensions. Lastly, a current sensing scheme was developed to provide real-time DEP-$I_D$ sensor data. A collection of samples was investigated to demonstrate the device's usefulness in sensing biological analytes, such as n-gene dsDNA, streptavidin-coated PS beads, and SARS-CoV2 VLPs. The demonstrated DEP-GFET technology provides a simple yet powerful way to concentrate target species at the surface of a GFET sensor without relying upon on diffusion-limited transport, which requires a long sensor response time.

**Supporting Information**

Supporting Information is available free of charge, and includes the following information: details of the graphene transfer, device fabrication, measurement setup, COMSOL simulations, and SEMs and videos of particle trapping.

**Statement of Financial Interests**

S.J.K. serves on the Scientific Advisory Board and holds and equity stake in GRIP Molecular Technologies. S.J.K. and S.-H.O. also have several patents related to this work and are entitled to standard royalties should licensing revenue be generated. These financial interests have been reviewed and managed by the University of Minnesota in accordance with its conflict-of-interest policies.

**Acknowledgements**

This work was supported primarily by Grip Molecular Technologies. Device fabrication was performed at the Minnesota Nanofabrication Center at the University of Minnesota, which receives partial support from the National Science Foundation (NSF) through the National Nanotechnology Coordinated Infrastructure (NNCI) under Award No. ECCS- 2025124. Portions of this work were also carried out in the University of Minnesota Characterization Facility, which receives capital equipment funding from the University of Minnesota MRSEC under award no. DMR-2011401.

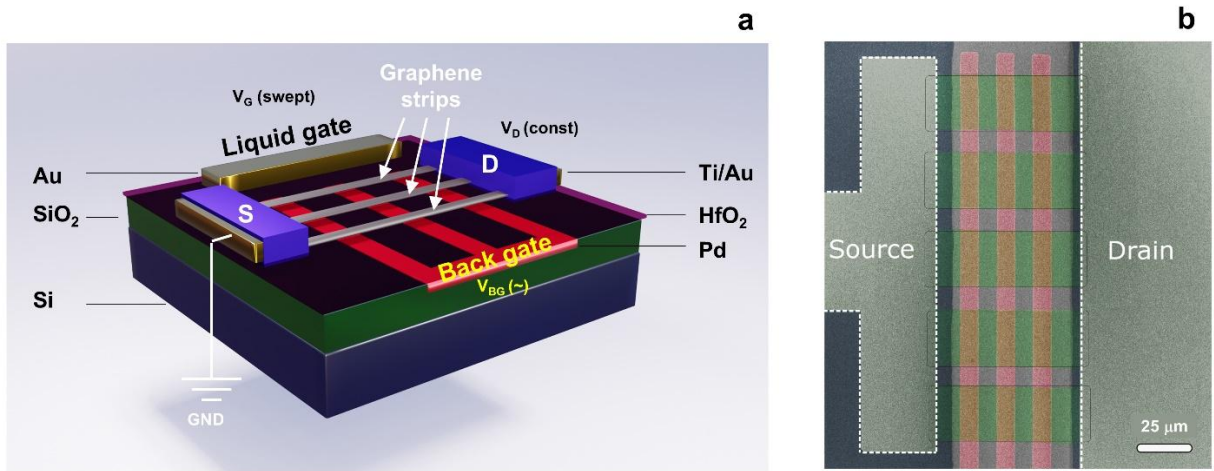

Figure 1. (a) Schematic diagram of DEP GFET device, and (b) color-enhanced top-down optical micrograph of completed DEP GFET device. The width of each graphene channel (green) is 25 μm while the designed width of each local back gate electrode (pink) is 7.5 μm. The total designed source-to-drain spacing is 70 μm.

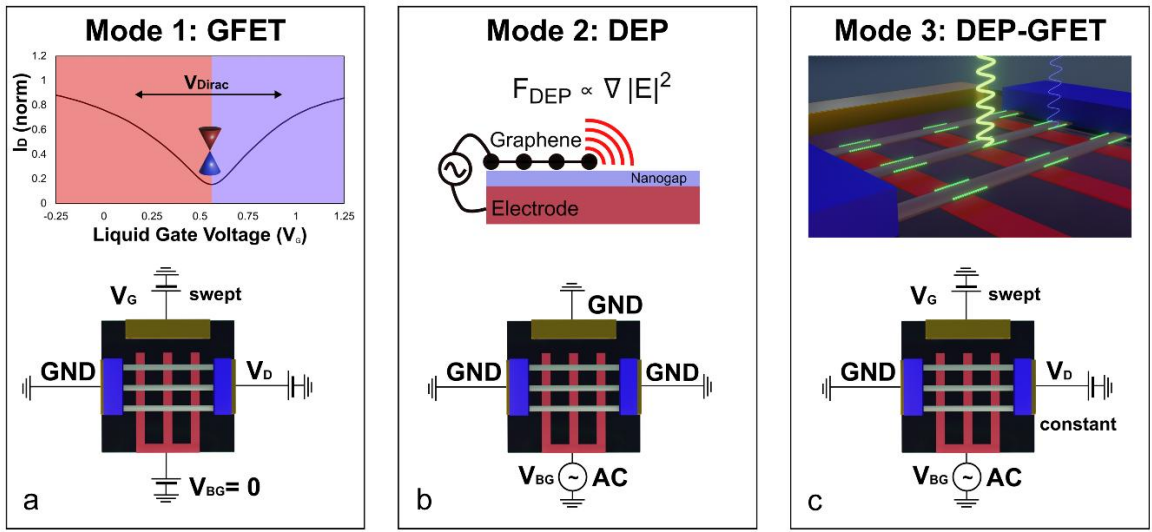

Figure 2. Diagram of different operating regimes for sensors. (a) GFET mode, (b) DEP mode, and (c) DEP-GFET mode.

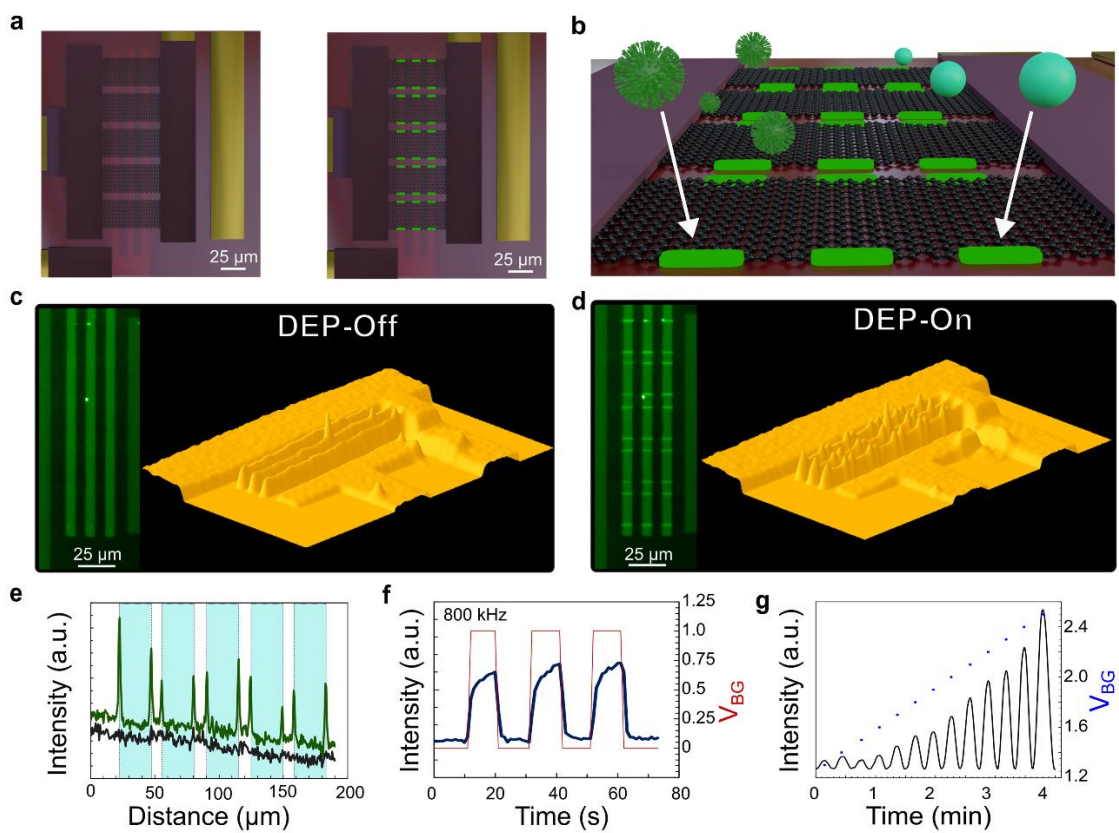

Figure 3. (a) Schematic depiction of DEP-off and DEP-on states. (b) Graphic depiction of intended mechanism of trapping of nanoparticles at graphene edges. (c)-(d) Fluorescent images and 3D intensity maps of polystyrene (PS) bead DEP trapping in DEP-GFET mode for both the (c) DEP-off and (d) DEP-on state. (e) Plot of fluorescence intensity vs. distance across the graphene sensor strip with DEP off (black) and DEP on (green). The data is extracted from results shown in Fig. 3c and 3d. Trapping is isolated to engineered trap sites located at the graphene-edge and local back gate intersection. (f) Tracking of fluorescence intensity (black) at the trap sites shows a reproducible baseline and trapping response for three sequential DEP pulses (red). (g) Filtered (Savitzky–Golay) data for trapping intensity (black line) at the trap sites shows a response to increasing DEP voltage ($V_{PP}$) (blue).

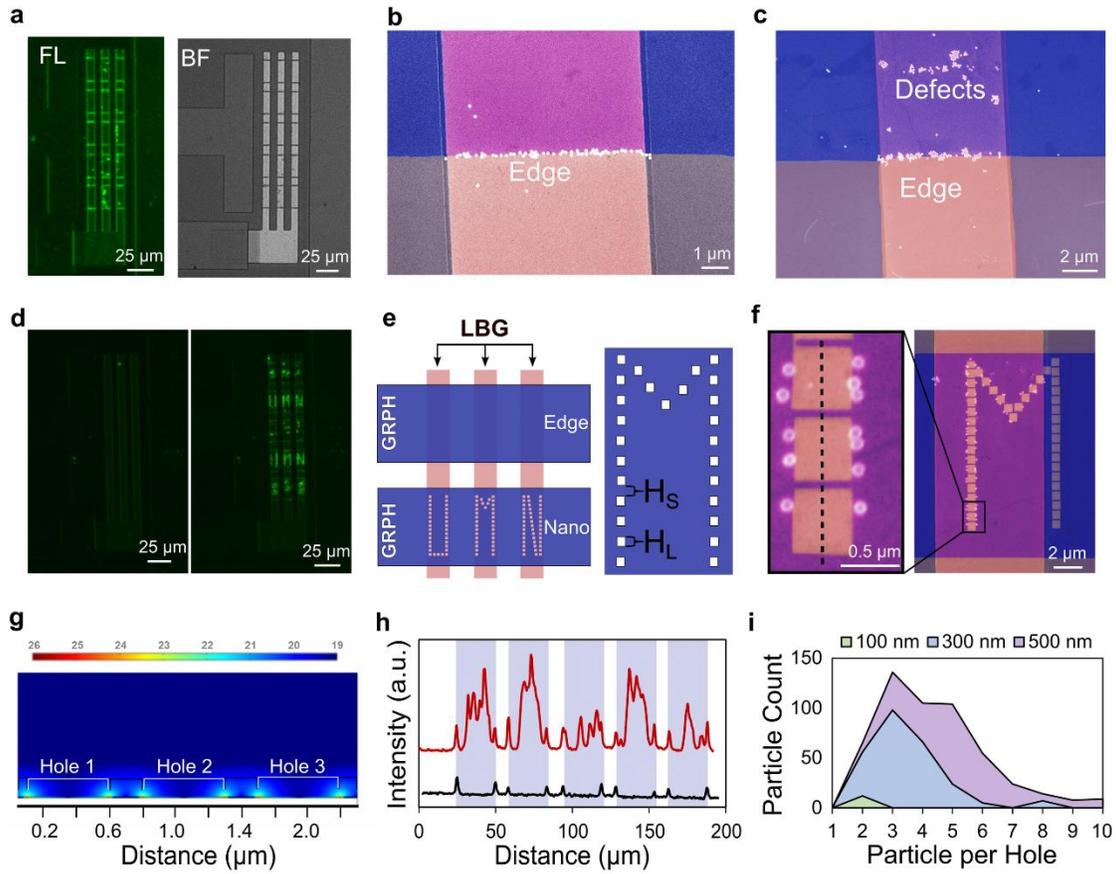

Figure 4. Designing nano-site trapping. (a) Fluorescence (FL) and bright field (BF) images of DEP trapping of AuNPs showing post-DEP particle immobilization above a threshold value of $V_{PP}$. (b)-(c) False color SEM images of trapping locations of (b) pristine and (c) defective graphene channels. (d) FL images before and after DEP showing AuNP trapping at electron-beam-lithography-defined nano-site. (e) Schematic diagram of nano-site design spelling out the letters U-M-N. (f) False color SEM images of AuNP trapping at broad nano-site edges. (g) COMSOL simulation of three neighboring nano-sites confirming the presence of strong electric field gradient. (h) FL line profile analysis of nano-site (red) trapping along the LBG shows a substantial increase in FL across the graphene channel width compared to edge trapping (black). (i) Comparison of FL immobilization for different nano-site dimensions $H_L$ = 100, 300 and 500 nm. The immobilization of particles per hole increases as $H_L$ increases, thus improving the immobilization efficiency.

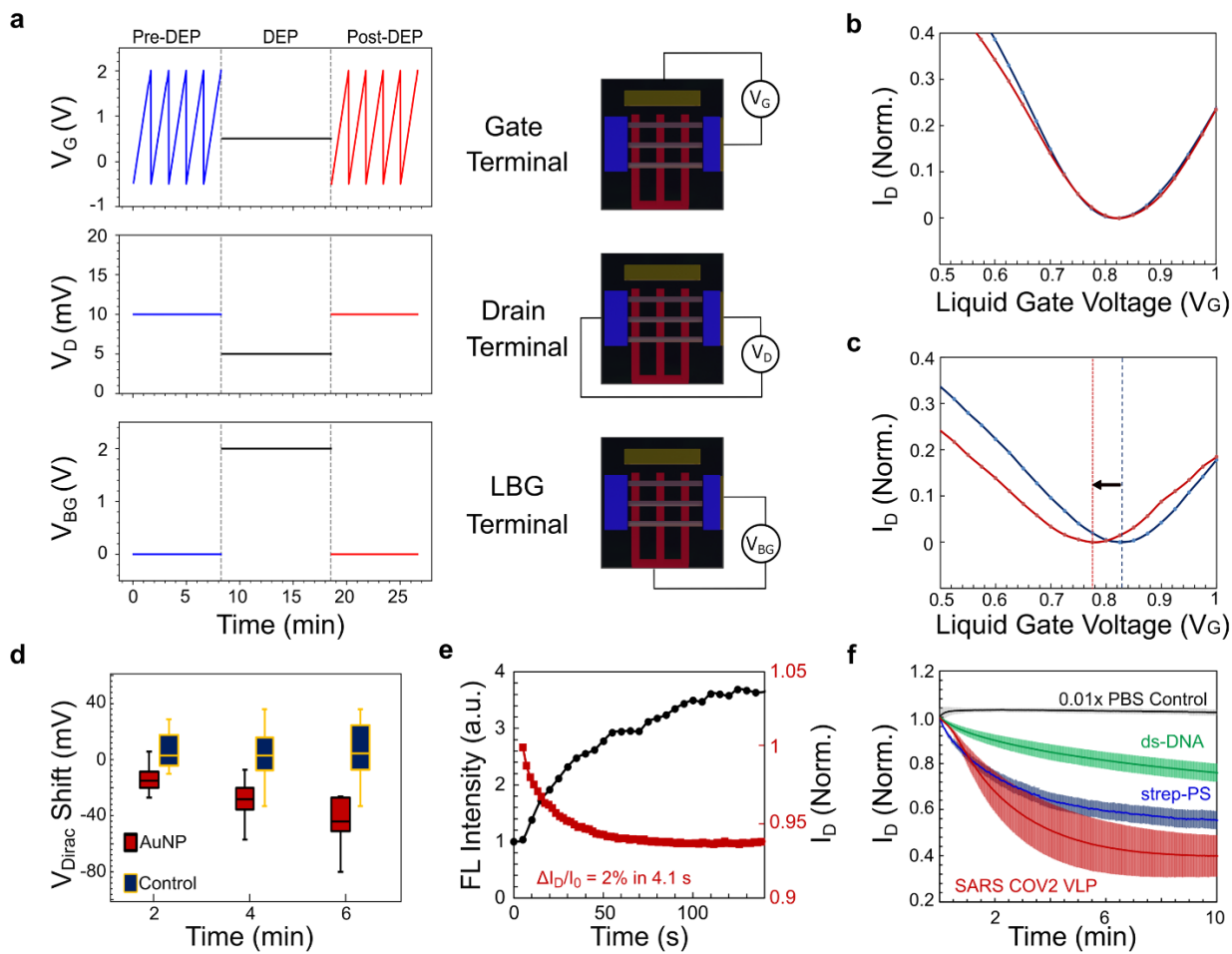

Figure 5. Sensing measurement protocol for DEP-GFETs. (a) Measurement sequence for Dirac voltage shift protocol, (b)-(c) results of Dirac voltage shift protocol, showing measurement after three sequential 2-minute DEP pulses for (b) a control sample, and (c) a sample of 100 pM AuNPs produced a -40 mV shift. (d) Dirac voltage shift after 2, 4 and 6 minutes of applied DEP for control and 100 pM AuNP sample. (e) Results for the same sample in (d), except measured using the continuous $I_D$-sensing mode, showing the current changes by 2% in only 4.1 s. (f) Results showing current shift for PBS control, ds-DNA (1 nM), streptavidin-coated spheres (1 pM), and SARS COV2 virus-like particles (VLPs) (5 fM).

# Supporting Information

# for

# Dielectrophoresis-Enhanced Graphene Field-Effect Transistors for Nano-Analyte Sensing

Nezhueyotl Izquierdo,[*] Ruixue Li,[*] Peter R. Christenson, Sang-Hyun Oh, and Steven J. Koester

*Department of Electrical and Computer Engineering, University of Minnesota, 200 Union Street SE, Minneapolis, MN 55455, United States*

[*]Equal contributors

## S1. Graphene Growth and Wet Transfer Procedure

Graphene growth on a Cu (25 μm x 25 mm x 25 mm, Alfa Aesar, 99.8% purity) foil was achieved on a home-built hot-wall chemical vapor deposition (CVD) system. Growth conditions were as follows: Gas flow rates: $H_2$ (0.105 SCCM) and $CH_4$ (21 SCCM), growth temperature: 1050 °C, growth pressure: 250 mTorr, growth time: 30 minutes. The growth was followed by a slow ramp-down to room temperature with $H_2$ (16 SCCM) flow. After growth, the foil was spin-coated on one side with PMMA dissolved at 4% by weight in chlorobenzene. An $O_2$ plasma was then used to etch off unwanted backside graphene. Afterward, the transfer was performed by soaking the graphene-coated Cu foil in ammonium persulfate (7 g/L) in $H_2O$ and then transferred onto the target substrate in DI $H_2O$.

## S2. DEP-GFET Device Fabrication

A schematic diagram of the device fabrication sequence is shown below in Figure S1. All lithography steps were completed using standard contact photolithography. The fabrication started using by patterning a Si/$SiO_2$ wafer and chemically recess-etching the $SiO_2$ followed by deposition of Ti/Pd (10/30 nm) as the local back gate (LBG) electrode. The LBG was arranged in multiple fingers to increase the number of DEP sites. The LG was patterned next, and is situated to the side of the sensor geometry, and consists of Ti/Au (10/90 nm). A 7-nm-thick $HfO_2$ was then deposited by atomic layer deposition (ALD) onto the LBG at 250 °C. The CVD-grown graphene was then transferred as described above and patterned into a series of stripes using optical lithography and $O_2$ plasma etching to form an array of parallel conducting channels. After forming openings through the $HfO_2$ (in extrinsic regions above the LBG), contacts to the graphene and LBG were made by evaporation and lifting off of Cr/Au/Ti (25/125/30 nm) by electron-beam evaporation. Finally, a passivation layer was deposited by evaporating a blanket 3-nm Al seed layer which was oxidized in air, followed by deposition of an $Al_2O_3$ layer (50 nm) by ALD at 180 °C. The $Al_2O_3$ above the active graphene channel region was etched away by patterning resist openings and performing chemical etching (Transetch-N, 55 °C) to remove the $Al_2O_3$, and then the resists was removed using a solvent strip process. This passivation layer helps to minimize leakage between the LG and the electrode leads.

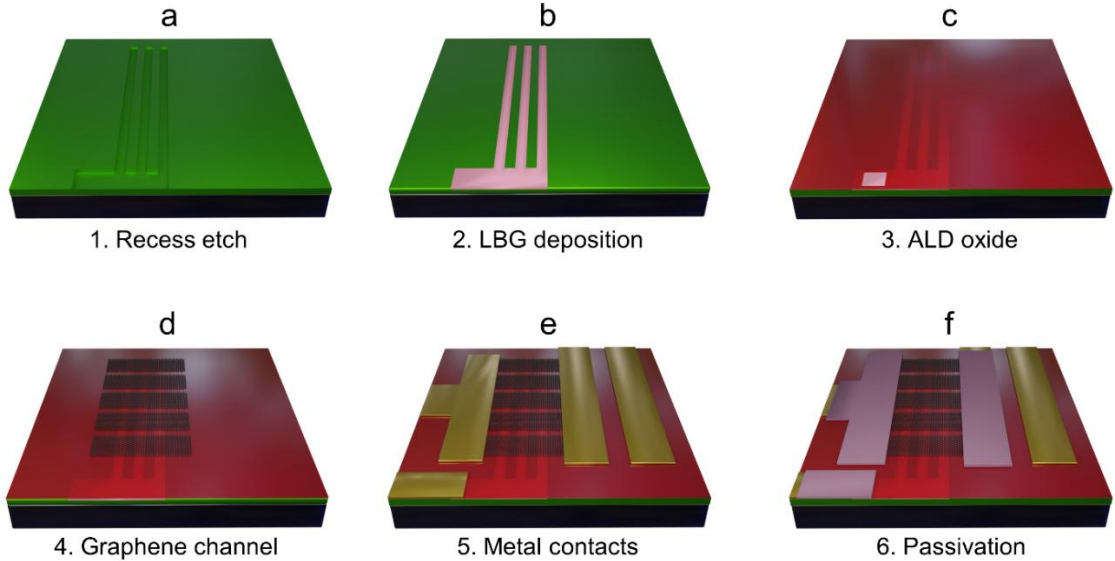

Figure S1. The fabrication process is detailed in six major steps. (a) Recess etch, (b) local backgate (LBG) deposition (Ti/Pd), (c) atomic layer deposition of $HfO_2$ as a LBG dielectric layer, (d) graphene channel layer transfer and patterning, (e) contact electrode deposition, and (f) source and drain electrode oxide passivation. Not shown is the final removal of the passivation layer in the active graphene channel region.

## S3. Fluorescence Microscopy and DEP-GFET Measurements

Fluorescence imaging was performed on a Nikon Eclipse LV100 microscope equipped with a CCD camera detector (Photometrics CoolSNAP HQ2). A laser-driven light source (Energetiq, model) coupled with an optical fiber cable illuminated the sample, through the fluorescence filter cube set ($\lambda_{ex}$ = 470 nm, $\lambda_{em}$ = 525 nm). Micro-Manager software collects raw CCD data at set time intervals. Post-capture images and video analysis were performed using ImageJ software. FL intensity, measured frame-by-frame, is plotted as a function of time. Extracted FL intensity can be monitored at precise locations, such as graphene-edge trap sites, through the creation of image masking.

The DEP and DEP-GFET sensing and trapping measurements were performed by depositing a 30 μL solution containing the active particles onto the chip surface. The liquid was contained by a silicone reservoir that was placed manually onto the chips surface. A micro glass slide was placed on top of the reservoir, creating an enclosed chamber in order to reduce the evaporation rate. The glass micro slide provides a flat surface for improved FL microscopic imaging. The FL microscope stage was modified to accommodate the four-contact probe micromanipulator. DEP and DEP-GFET modes were tested while measuring the FL response. Electrical bias to the LG and drain electrode were controlled with two Keithley 2400 Source Measure Units (SMU), while the DEP signal was applied to the LBG using an Agilent 33600A arbitrary waveform generator. In all measurements, the source contact remained at ground potential. A diagram of the test setup is shown in Figure S2.

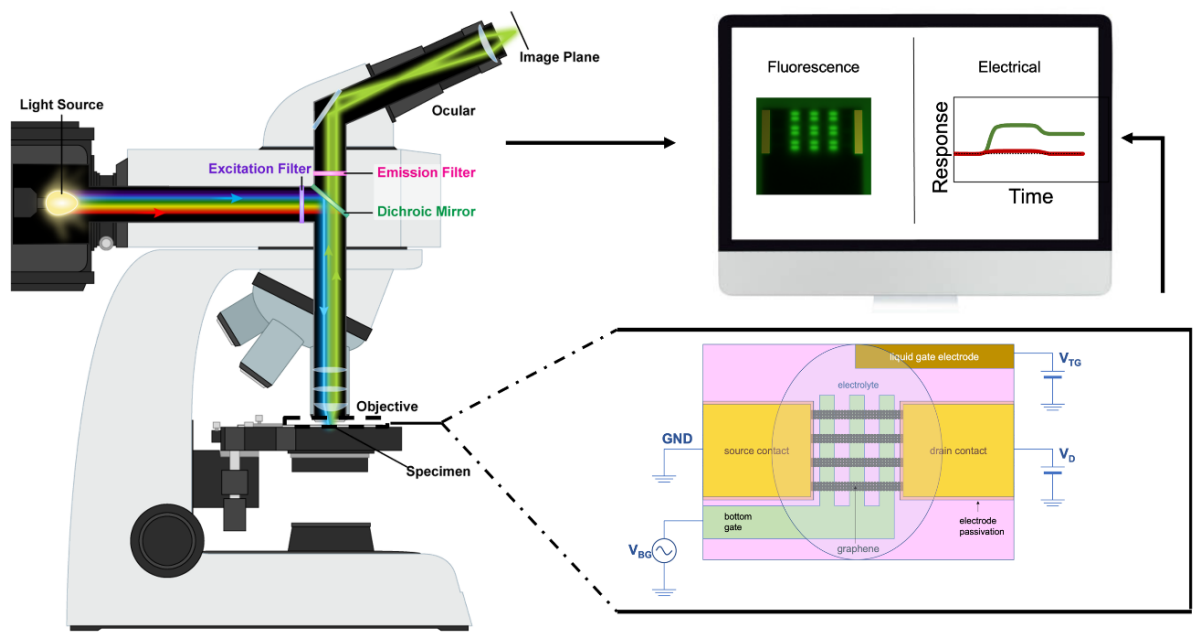

Figure S2. Schematic diagram of measurement setup, showing simultaneously measurement using optical microscope for fluorescence measurements, combined with simultaneous electrical readout using GFET measurements.

## S4. DEP-GFET Modes of Operation

In the "GFET" sensing mode, DEP is not used and the device operates as a conventional GFET sensor. In this configuration, the sensor tracks the Dirac point, $V_{Dirac}$, and/or the drain current, $I_D$. Experimentally, when $V_{Dirac}$ is extracted, the source and LBG are grounded, a DC voltage, $V_D$, is applied to the drain, and the LG voltage, $V_G$, is slowly swept from -2 to +2V and back to -2 V in order to map out the location of $V_{Dirac}$. This is repeated as the sensing fluid is applied to the device and the shift in $V_{Dirac}$ is used as a means to detect changes in the surface charge in the graphene sensor channel. Alternatively, one can also measure the GFET by applying a fixed voltage to the LG, and then monitoring $I_D$ continuously, which provides a real-time response, though with less information about the nature of the analyte interaction. Either method is acceptable for sensing shifts in the surface charge in the graphene channel.

In the second "DEP" mode, the GFET is not used, and only fluorescence microscopy is used to identify trapping at the graphene edges. Here, an AC bias, $V_{pp}$, is applied to the LBG, and all other electrodes are grounded. This mode is particularly useful to optimize the trapping parameters for a particular device before being operated as an electrical sensor, since rapid feedback on the effectiveness of the different bias and frequency conditions can be obtained. A final sensing mode is the "DEP-GFET" configuration. In this mode, a DC bias, $V_D$, is applied between source and drain, and the AC bias for DEP, $V_{pp}$, is applied to the LBG, while the liquid gate bias, $V_G$, used to modulate the conductivity of the graphene to identify the Dirac voltage, $V_{Dirac}$, or monitor $I_D$.

## S5. COMSOL Simulations

Modeling of the electric field gradient was performed in the COMSOL Multiphysics 2D electrostatic module. The LBG electrode was modeled as a perfect conductor. Graphene was approximated as a perfect conductor with a finite thickness (0.34 nm). A voltage of 1 or 2 V was applied to the graphene channel and the LBG electrode was grounded. The dielectric constants of $H_2O$ and $HfO_2$ (7 nm) were set to 80 and 13, respectively.

## S6. Overview of DEP Trapping Fundamentals

A DEP force, $F_{DEP}$, occurs when the dipole of a particle interacts with a non-uniform field in a liquid buffer solution or medium. The following equation[S1] describes the time-averaged DEP force on a spherical particle:

$$\vec{F}_{DEP}(\omega) = \pi\varepsilon_m R^3 \cdot Re\left(\frac{\varepsilon_p^*(\omega)-\varepsilon_m^*(\omega)}{\varepsilon_p^*(\omega)+2\varepsilon_m^*(\omega)}\right)\nabla|E|^2, \tag{S1}$$

where the complex permittivity of the medium, $\varepsilon_m$, particle radius, $R$, and gradient of the electric-field squared, $\nabla|E|^2$, are considered. $F_{DEP}$ scales in proportion to $\nabla|E|^2$, which is affected by the electrode radius of curvature. According to previous work, the regions where the edge of an $sp^2$ planar graphene (thickness = 0.3 nm)[S2] sheet overlaps the LBG electrode edge can achieve 10× higher gradient forces compared to a 20-nm-thick metal electrode.[S3] Whether or not $F_{DEP}$ is positive or negative depends upon another important parameter, the frequency-dependent

Clausius-Mossotti (CM) factor, $f_{CM}(\omega)$, is used to describe the polarity of the DEP force.[S4] The CM factor is given by

$$f_{CM}(\omega) = \frac{\varepsilon_p^*(\omega) - \varepsilon_m^*(\omega)}{\varepsilon_p^*(\omega) + 2\varepsilon_m^*(\omega)}. \tag{S2}$$

Here, the crossover frequency occurs when $f_{CM}(\omega) = 0$ and refers to the AC frequency at which a transition between the positive DEP (pDEP, attractive = $(f_{CM}(\omega)) > 0$) and negative DEP (nDEP, repulsive = $(f_{CM}(\omega)) < 0$)) force occurs. This switch-over occurs when the polarizability of the particle equals the polarizability of the liquid medium. When operated in DEP-only mode, the liquid is left floating, the source and drain are shorted, and an AC bias is applied between the LBG and the source/drain electrodes.

## S7. Video of Polystyrene (PS) Bead Trapping / Detrapping

S. Video 1 shows the result of sequential DEP (on-off) pulses measured by monitoring the mean FL intensity at all trap sites.[S5] The device was operated at a frequency of $f = 800$ kHz. In the OFF trapping condition, $V_{pp} = 0$, while in the ON trapping condition, $V_{pp} = 1.5$ V. The polystyrene (PS) beads had roughly 42 nm diameter. The results indicate the device is capable of efficiently trapping and releasing the particles.

## S8. Video of Effect of LBG Voltage on Polystyrene Bead Trapping

S. Video 2 shows the effect of increasing $V_{PP}$ on the FL intensity.[S6] The device was tested at various $V_{pp}$ levels (1.5 $V_{pp}$, 2 $V_{pp}$, and 2.5 $V_{pp}$) while maintaining a frequency, $f = 800$ kHz in the ON state, and $f = 8$ MHz in the OFF state. The results clearly show that trapping intensity increased with higher $V_{pp}$ values. This indicates that the device's trapping efficiency is dependent on $V_{pp}$ (intensity) and demonstrates minimal permanent trapping.

## S9. Video of AuNP Trapping at Nano Sites

S. Video 3 shows FL trapping of AuNPs in nano-sites formed in the graphene channel.[S7] In the case of the chosen nano-site pattern, DEP-trapping is shaped into a distinct outline of the letters U-M-N. The trapping mode was operated under ~1 $V_{pp}$ and $f \sim 1$ MHz.

## S10. Nano-site Fabrication and SEM Analysis of AuNP Trapping

To evaluate enhanced trapping efficiency, electron-beam lithography was used to create the nano-sites traps within the graphene channel. The nano-sites were created by patterning poly(methl methacrylate) (PMMA) on the completed devices from Figure 1 in the main manuscript, and creating a series of openings in the resist of various sizes and shapes, followed by a short 10 second $O_2$ plasma step, and then a solvent resist stirp. The $O_2$ plasma is sufficient to remove the underlying graphene, but leaves the $HfO_2$ relatively undamaged.

We investigated the effect of the nano-site dimensions on the trapping. Here, we considered the hole spacing, $H_S$, and hole length, $H_L$. We observed that the nano-site dimensions affect the DEP trapping. Increasing $H_L$ was found to increase the number of trapped AuNPs, as expected, resulting in an overall increase in the FL intensity. To quantify this effect, we determined the permanent trapping occupancy (PTO), defined as:

$$PTO = \frac{\# \ of \ filled \ trap \ sites}{total \ trap \ sites} \ x \ 100, \quad (3)$$

where PTO values were measured for $H_L$ = 100, 300, and 500 nm. Here, the # of filled trap sites was defined as the number of nano-sites with at least one trapped AuNP. The resulting values were PTO = 5.7%, 90%, and 97.6%, respectively. Furthermore, the number of particles per hole increased as $H_L$ increased. Figure S3 shows SEM micrographs and statistical analysis of the dependence of AuNP trapping efficiency as a function of nano-site dimensions.

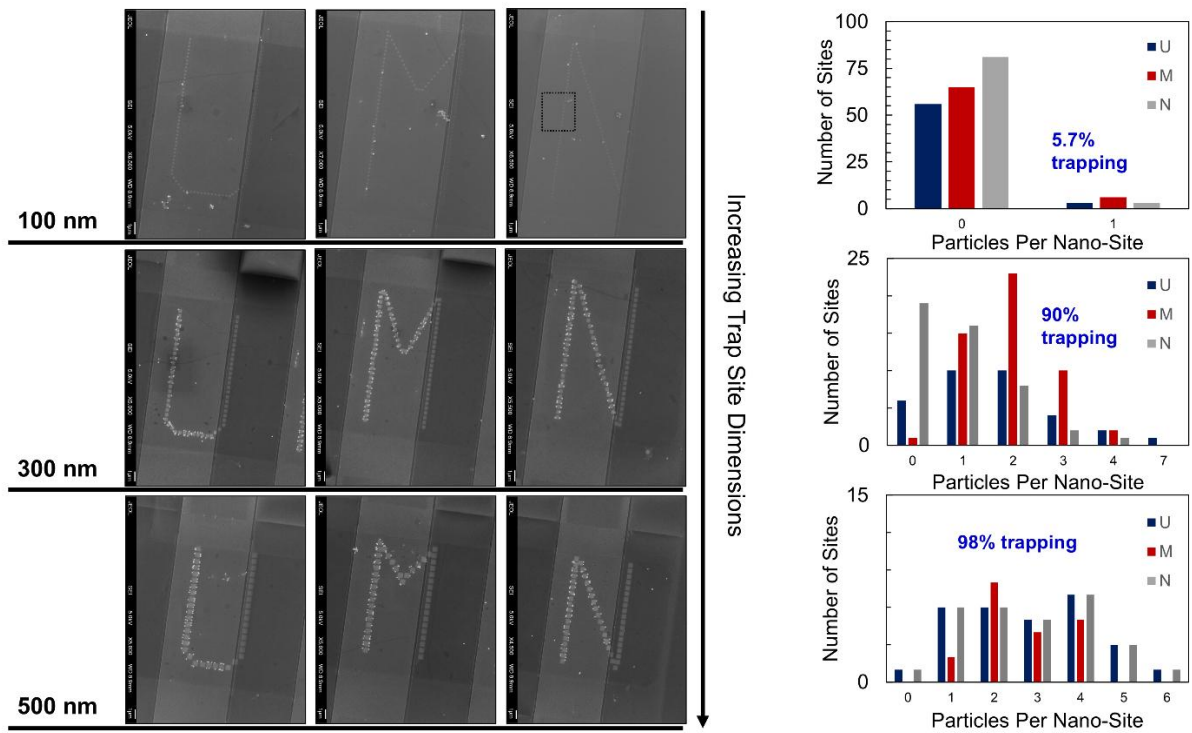

Figure S3. Left: SEM micrographs of S-E-M patterned nano-sites of different sizes. Each nano-site is a square with edge length of 100 nm, 300 nm or 500 nm. Right: Number of particles trapped at each letter of the nano-site along with the trapping efficiency, defined as the percentage of nano-sites with at least one AuNP trapped at the edge. The trapping efficiency is greatly suppressed with the nano-site size is equal to, or smaller than, the AuNP diameter.

## S11. DEP-GFET Electrical Characterization

Figure S4 shows the pre-sensing measurement results on the DEP-GFETs to understand how application of the DEP bias affects device performance.

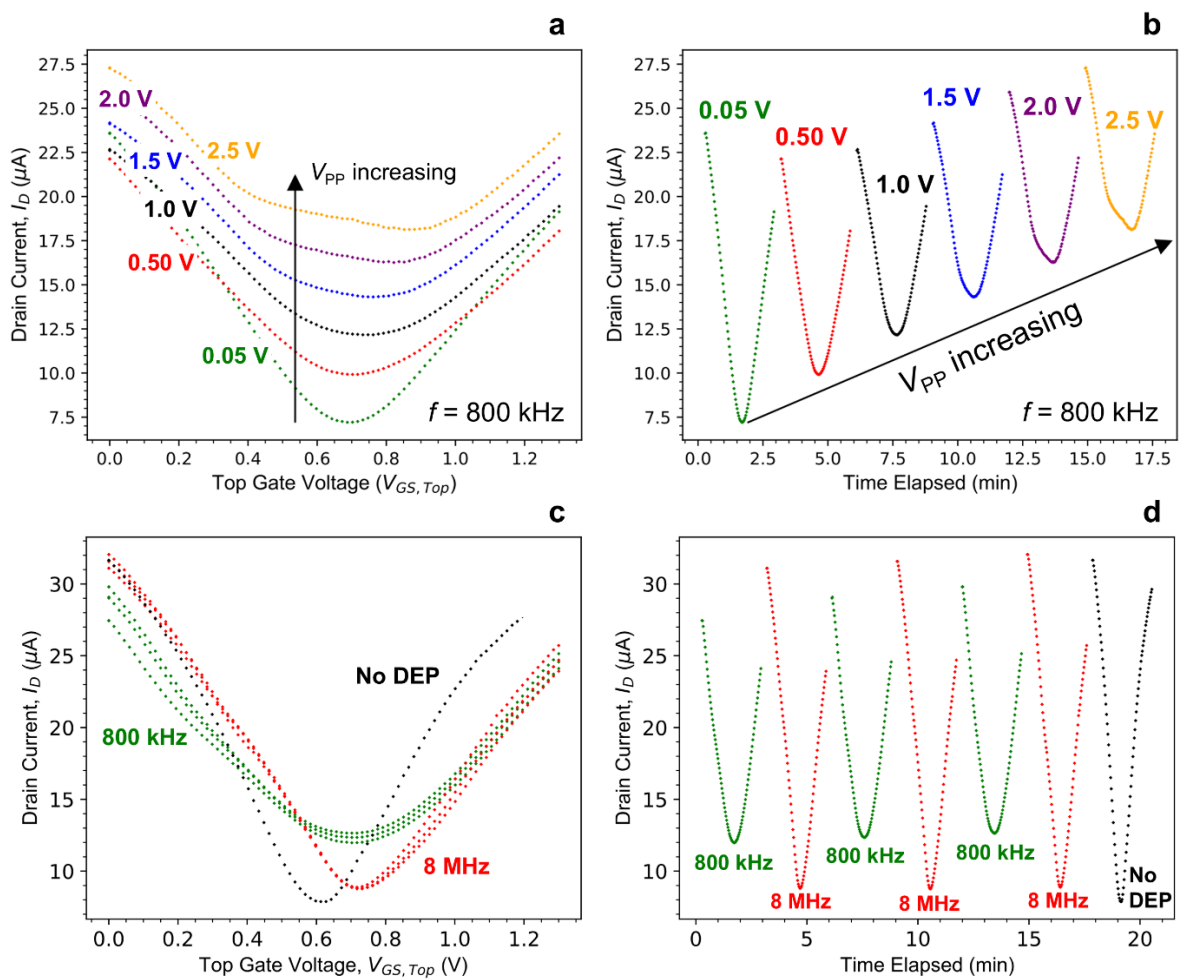

Figure S4. (a) $I_D$-$V_G$ plot for a typical DEP-GFET device showing the effect of increasing $V_{pp}$ on the GFET characteristics. The results show that a smearing effect occurs, but that the Dirac point remains observable, even up to $V_{pp} = 2.5$ V. (b) Plot showing $I_D$ vs. time for same data in (a). (c) $I_D$-$V_G$ plot showing effect of DEP frequency on GFET characteristics. Here $V_{pp} = 0.5$ V. The results show that lower frequencies have a strong effect on the characteristics due to RC time delays which reduce the actual voltage that reaches the active device. (d) Plot showing $I_D$ vs. time for same data in (c).

## S12. DEP-GFET AuNP Sensing Details

For sensing of AuNPs using DEP-GFETs, the following procedure was used. Step 1: a pre-DEP baseline transfer curve measurement was taken before the application of DEP, where an initial $V_{Dirac}$ position is established to evaluate relative shift post-DEP. In our experiments, $V_G$ was swept, while $V_D$ was held constant at 50 mV. A DEP force was not applied and only a minimum DC bias of 1 mV was applied to the LGB. Step 2: a DEP force was applied while simultaneously measuring current, providing a method to measure the direct-electrical response to DEP trapping. In this scheme, the LG and drain voltage were held constant at $V_G = 0.5$ V and $V_D = 50$ mV. The DEP parameters, such as the $V_{PP}$ and AC bias frequency, were adjusted for optimal trapping and applied depending on the target particle. Step 3: a post-DEP response transfer curve measurement was performed to track changes in $V_{Dirac}$, relative to the baseline position. Here, $V_G$ was swept using the same parameters as step 1, while $V_{DS}$ was held constant at 50 mV. Once again, $F_{DEP}$ was not applied, and only a 1 mV DC bias was applied to the LBG. In summary, steps 1 and 3 track the $V_{Dirac}$ shift, as an indirect DEP sensing method, responsive only to immobilized particles. In certain DEP conditions, such as low voltage DEP, immobilization is not observed. Therefore, a direct DEP-$I_D$ sensing method, such as afforded by step two, is a desirable measurement. Our measurements in DEP-GFET mode used fluorophore-decorated-AuNPs to evaluate the sensing protocol. The active (100 pM, AuNP in 0.01X PBS) and control (0.01X PBS) samples were tested as described above using precisely the same sensing protocol.

## S13. DEP-GFET Biomolecule Sensing Details

For the double-stranded DNA (ds-DNA) trapping, 4k base-pair n-gene SARS-CoV2 ds-DNA at 1 nM concentration was tested in 0.01X PBS. The DEP frequency was $f = 800$ kHz and the bias applied to the LBG was 2 $V_{PP}$. In another experiment, PS-beads coated with streptavidin protein were investigated. We used 198-nm latex-coated streptavidin (1 pM in 0.01X PBS). For these measurements, the DEP frequency was 1 MHz and the bias applied to the LBG was 2 $V_{PP}$. In another experiment, we tested the readout of virus-like particles (VLPs) (GenTarget Inc.) coated with a SARS-CoV2 spike protein (s protein) exterior such that they are capable of binding to S protein binding receptors (e.g., ACE2). The VLP diameter was ~90 nm to resemble the SARS-CoV2 viral particle dimensions. Sensing was performed at a concentration of 5 fM in 0.5X PBS, at a frequency of $f = 800$ kHz and 2 $V_{PP}$. In these measurements, the VLP was engineered to express green fluorescent protein (GFP) in the particle core as virion genomic material, and an FL image of VLP trapping is shown in Figure S5, along with a line scan along the length of each LBG electrode, to indicate that the VLPs are isolated to the DEP trapping region and absent in the electrode region, as expected. In the FL imaging, the VLP concentration was 100 fM in 1X PBS.

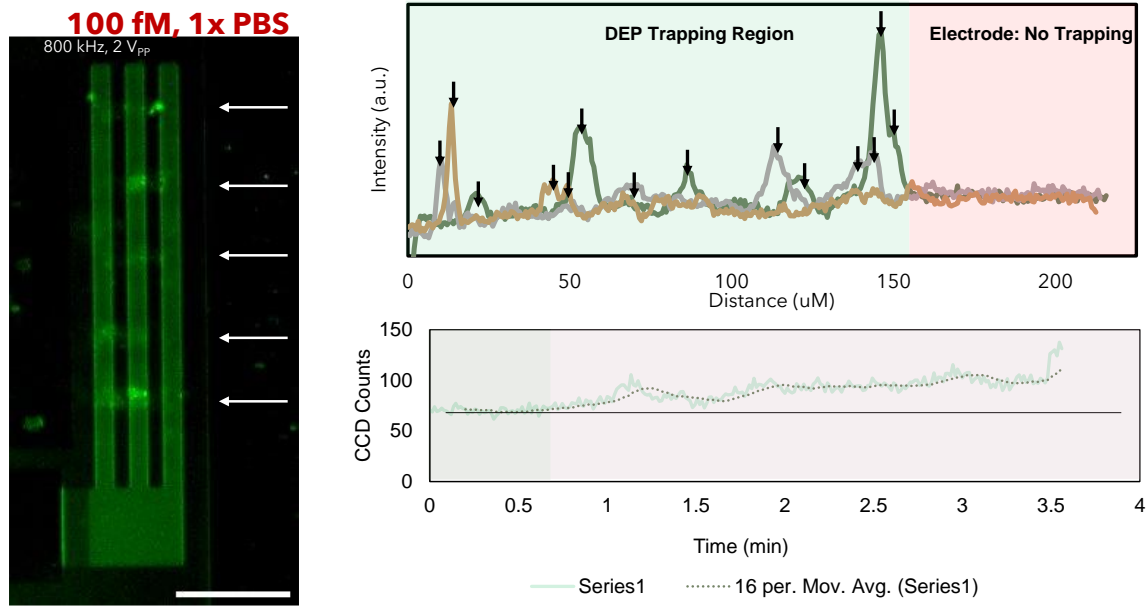

Figure S5. Trapping of virus-like particles (VLPs) using DEP-mode and fluorescence imaging. Left: Fluorescence image of trapping. Right: Spatial and time series line scans.